\documentclass[a4paper,11pt]{article}
\pdfoutput=1 

\usepackage{jheppub} 

\usepackage[T1]{fontenc} 
\usepackage{appendix}
\usepackage{comment}

\title{\boldmath Spinor-Helicity Formalism for Massless Fields in AdS$_4$ III: Contact Four-Point Amplitudes}

\author[a]{Balakrishnan Nagaraj}
\author[b,c]{and Dmitry Ponomarev}

\affiliation[a]{George P. and Cynthia W. Mitchell Institute for Fundamental Physics and Astronomy,\\
 Texas A\&M University, University Drive,  College Station, TX 77843, USA}
\affiliation[b]{Institute for Theoretical and Mathematical Physics,\\
Lomonosov Moscow State University, Lomonosovsky avenue, Moscow, 119991, Russia}
\affiliation[c]{I.E. Tamm Theory Department, Lebedev Physical Institute,\\
 Leninsky avenue, Moscow, 119991, Russia}

\emailAdd{nbala1090@gmail.com}
\emailAdd{ponomarev@lpi.ru}

\abstract{We study contact four-point amplitudes in the spinor-helicity formalism in anti-de Sitter space. We find that these amplitudes can be brought to an especially simple form, which we call canonical. Next, we classify consistent contact amplitudes by requiring correct transformation properties with respect to the AdS isometry algebra. Finally, we establish a connection between the canonical form of AdS amplitudes and scalar multi-trace conformal primaries in flat space.}

\begin{document} 
\maketitle
\flushbottom

\section{Introduction}

The spinor-helicity formalism is a technique that allows to evaluate scattering amplitudes in massless theories in four-dimensional flat space efficiently. It also brings them to a very simple and compact form. Thanks to these virtues, the spinor-helicity formalism has become an important part of modern amplitude methods; for review see \cite{Dixon:1996wi,Elvang:2013cua,Dixon:2013uaa}. Aiming to achieve similar simplifications for computations that appear in the context of holography and higher-spin theories, we recently extended the spinor-helicity formalism to four-dimensional anti-de Sitter space \cite{Nagaraj:2018nxq,Nagaraj:2019zmk}\footnote{Our approach is closely related to the twistor-space formalism applied to amplitudes in AdS space; see e.g. \cite{Adamo:2012nn,Skinner:2013xp}. Moreover, there is an alternative spinor-helicity formalism for AdS${}_4$ suggested in \cite{Maldacena:2011nz}. For a more comprehensive review of the relation of our formalism to other approaches, see \cite{Nagaraj:2019zmk}.}. In particular, we constructed the AdS${}_4$ counterpart of the plane-wave solutions for spinning fields and then employed them to evaluate some spinning three-point amplitudes. We also classified three-point spinor-helicity amplitudes of spinning fields by requiring correct transformation properties with respect to the AdS${}_4$ isometry algebra.

In the present paper we initiate the analysis of  spinor-helicity amplitudes in AdS${}_4$ beyond three points. Unlike three-point amplitudes, which are fixed by symmetries up to a coupling constant, symmetries constrain four-point amplitudes up to a function of two variables. Hence, it makes sense to talk about their analytic structure and study its relation to the type of the diagram, the amplitude originates from. Understanding of the analytic structure of amplitudes is a key step towards the development of the on-shell methods, which proved to be an efficient approach to computing them\footnote{See, for example, \cite{smatrix} for a classical review on the flat space S-matrix and its analytic structure.  Analytic structure of AdS Witten diagrams has been studied in various representations and in many cases it is well understood; see \cite{Liu:1998ty,Freedman:1998bj,Liu:1998th,DHoker:1998ecp,Heemskerk:2009pn,Penedones:2010ue,Fitzpatrick:2011ia,Fitzpatrick:2011hu,Fitzpatrick:2011dm,Aharony:2016dwx,Caron-Huot:2017vep,Alday:2017vkk,Alday:2017gde,Yuan:2018qva,Arkani-Hamed:2018kmz,Ponomarev:2019ofr,Meltzer:2019nbs} for a far from complete list of references. 
These achievements were used, in particular, for bootstrapping holographic amplitudes in type-IIB supergravity \cite{Rastelli:2016nze} and for computing  the associated  loop corrections from the dispersion relations \cite{Alday:2017xua,Aprile:2017bgs}.}.

Below we will focus on contact four-point amplitudes of scalar fields involving arbitrary number of derivatives. In flat space the result of this computation is well-known. Namely, amplitudes are given by polynomials in the Mandelstam variables with the degree of a polynomial being equal to the half of the number of derivatives. Moreover, the amplitude comes supplemented with the momentum-conserving delta function as a factor, which entails familiar equivalence relations for the Mandelstam variables. We would like to obtain an analogous statement for the spinor-helicity amplitudes in AdS space.

The key difference of the AdS analysis is that translation invariance is absent. This implies that the action contains manifest dependence on the space-time coordinates and, as a result, amplitudes contain derivatives of the momentum-conserving delta function. This, in turn, entails that the standard equivalence relations on the Mandelstam variables no longer  hold in AdS space. Instead,  AdS space amplitudes satisfy more complicated equivalence relations. We explore these relations and find that amplitudes in AdS space can be brought to a certain particularly simple form, which we call canonical; see (\ref{20mar5}), (\ref{20mar6}) for the definition. 

We then proceed to the classification of consistent contact AdS four-point amplitudes. To this end, we consider  most general amplitude in the canonical form and require that it transforms appropriately with respect to the AdS space isometries. Solving the ensuing constraints, we find that, as in flat space, consistent amplitudes in AdS space can be labelled by polynomials of two variables. We then provide an alternative perspective on these results, which is based on the conformal equivalence of the flat and AdS spaces.

The paper is organized as follows. In the next section we briefly review the necessary results from the spinor-helicity formalism in AdS${}_4$. In section \ref{AUPW} we compute few lower-derivative four-point amplitudes from the action. We show that they can be brought to the canonical form, which we define. Next, in section \ref{AFS} we make a general ansatz for the four-point amplitude in the canonical form and then require that it transforms properly under AdS space isometries. By solving the resulting constraints,  we establish a classification of the contact four-point amplitudes. In section \ref{sec5:weyl} we motivate the canonical form of the AdS amplitudes from the conformal/Weyl symmetry.
 In section \ref{sec6:conc} we give our conclusions. A number of appendices contains our notations and some technical details.

\section{Preliminaries}
In this section, we collect some useful results on the spinor-helicity formalism in AdS${}_4$. For more comprehensive review and references, we refer the reader to the original papers \cite{Nagaraj:2018nxq,Nagaraj:2019zmk}. Our conventions are collected in appendix \ref{appa:conventions}.

The massless representations of the AdS$_4$ isometry algebra $so(3,2)$ can be realized as
\begin{equation}
\label{isometry}
\begin{split}
\mathcal{J}_{\alpha \beta}&=i \left(\lambda_{\alpha}\frac{\partial}{\partial\lambda^{\beta}}+\lambda_{\beta}\frac{\partial}{\partial\lambda^{\alpha}}\right),\\
\mathcal{\bar{J}}_{\dot{\alpha}\dot{\beta}}&=i\left(\bar{\lambda}_{\dot{\alpha}}\frac{\partial}{\partial\bar{\lambda}^{\dot{\beta}}}+\bar{\lambda}_{\dot{\beta}}\frac{\partial}{\partial\bar{\lambda}^{\dot{\alpha}}}\right),\\
\mathcal{P}_{\alpha\dot{\alpha}}&=\lambda_{\alpha}\bar{\lambda}_{\dot{\alpha}}-\frac{1}{R^2} \frac{\partial}{\partial\lambda^{\alpha}}\frac{\partial}{\partial\bar{\lambda}^{\dot{\alpha}}},\\
\end{split}
\end{equation}
where $R$ is the AdS space radius. By taking the flat space limit $R\to\infty$, we recover the usual Poincar\'e algebra in the spinor form. This representation of the AdS isometry algebra is often referred to as the twisted adjoint representation and is used extensively in the higher-spin literature \cite{Bekaert:2005vh}.

We will be working with AdS$_4$ in the stereographic coordinates, as they make the Lorentz symmetry manifest. 
The metric is given by
\begin{equation} \label{MT}
    ds^2=G^{-2}\eta_{\mu\nu}dx^{\mu}dx^{\nu},
\end{equation}
with the conformal factor 
\begin{equation}
\label{24oct19}
G\equiv 1-\frac{x^2}{4R^2}.
\end{equation}
Covariant derivatives are defined by 
\begin{equation}
\label{23oct19}
\nabla_\nu v_{\lambda}\equiv \partial_\nu v_{\lambda}-\Gamma_{\nu|}{}^\rho{}_{\lambda}v_\rho,\qquad
\nabla_\nu v^{\lambda}\equiv \partial_\nu v^\lambda +\Gamma_{\nu|}{}^\lambda{}_{\rho}v^\rho
\end{equation}
and the Christoffel symbols read
\begin{equation}
\label{23oct20}
\Gamma_{\nu|}{}^{\rho}{}_{\lambda}={\left(2R^2G\right)^{-1}}
\left(x_\nu \delta_\lambda^\rho +x_\lambda \delta_\nu^\rho - x^\rho \eta_{\nu\lambda} \right).
\end{equation}

The AdS space counterpart of the flat plane-wave solutions in the scalar case are given by\footnote{These solutions can be truncated to other patches without breaking AdS covariance. Here, we will only consider solutions (\ref{SPW}), which are supported on the global AdS patch.}
\begin{equation} \label{SPW}
    \phi=G e^{ipx} \qquad \text{with} \qquad p^2=0.
\end{equation}
As usual, by exploiting the vector-spinor dictionary, a light-like momentum $p$ can be factorized into a product of two $sl(2,\mathbb{C})$ spinors
\begin{equation}
\label{22oct2}
p_a=-\frac{1}{2}(\sigma_a)^{\dot\alpha\alpha}\lambda_\alpha\bar\lambda_{\dot\alpha},
\end{equation}
where $\sigma_a$ are the Pauli matrices. It is straightforward to check that (\ref{SPW}) satisfies the zero-mass free wave equation in AdS 
\begin{equation}
\label{29oct7}
\left(\Box + \frac{2}{R^2} \right)\phi\approx 0.
\end{equation}
 By taking the limit $R\to\infty$ of (\ref{SPW}), we reproduce the familiar flat-space plane waves. In the next section AdS plane waves (\ref{SPW}) will be used to calculate some lower-derivative four-point amplitudes.

\section{Amplitudes from Plane Waves} \label{AUPW}
In anti-de Sitter space tree-level amplitudes can be defined as the on-shell action evaluated on the solutions to the free equations of motion. In this section, we will use this definition and solutions (\ref{SPW}) to evaluate some simple lower-derivative contact four-point amplitudes for scalar fields.

As a warm up exercise, we first consider a no-derivative four-point vertex
\begin{equation} \label{vertex1}
    S_4=\int d^4 x\sqrt{-g}\phi^4.
\end{equation}
By plugging in  (\ref{SPW}) and the background metric (\ref{MT}) one can see that conformal factors cancel out. Evaluating the resulting integral, we obtain
\begin{equation}
\label{trivial}
    A_4=\int d^4x e^{i(p_1+p_2+p_3+p_4)x} = (2\pi)^4 \delta^{(4)}(p),
\end{equation}
where $p$ is the total momentum.
This result is identical to what one gets in flat space, because vertex (\ref{vertex1}) is conformally invariant.

Next, we proceed to a vertex with two derivatives
\begin{equation}
\label{20mar7}
    S_4=\int d^4 x\sqrt{-g}(\nabla^{\mu}\phi_1)(\nabla_{\mu}\phi_2)\phi_3\phi_4.
\end{equation}
 By substituting the plane-wave solutions and the metric we find
\begin{equation} \label{amp01}
        S_4=\int d^4x\left[\frac{x^2}{4R^4}-\frac{i}{2R^2}(p_1+p_2)x G-p_1p_2G^2\right]e^{ipx}.
\end{equation}
Evaluating the Fourier integral, we arrive at the amplitude in the form
\begin{equation} \label{amp1}
        A_4=(2\pi)^4\left[-\frac{\Box_p}{4R^4}-\frac{1}{2R^2}(p_1+p_2)\cdot\frac{\partial}{\partial p} \tilde G-p_1p_2\tilde G^2\right]\delta^{(4)}(p),
\end{equation}
where 
\begin{equation}
\label{20mar1}
\tilde G\equiv 1+\frac{\Box_p}{4R^2}
\end{equation}
is the Fourier transform of factor $G$.

The right hand side of (\ref{amp1}) already gives a valid formula for the amplitude we are computing. So, in principle, one can stop at this point and move on to other examples. However, it is important to note that (\ref{amp1}) is not the only representation for a given amplitude. For example, by using that
\begin{equation}
\label{20mar2}
p_\nu \tilde G^2 \delta^{(4)}(p)=-\frac{1}{R^2}\frac{\partial}{\partial p^\nu} \tilde G \delta^{(4)}(p),
\end{equation}
we can rewrite the second term in (\ref{amp1}) as
\begin{equation}
\label{20mar3}
 -\frac{1}{2R^2}(p_1+p_2)\cdot\frac{\partial}{\partial p} \tilde G\delta^{(4)}(p)=
 \frac{1}{2}(p_1+p_2) \cdot p \;\tilde G^2 \delta^{(4)}(p).
\end{equation}
We will often refer to transformations of this type as integration by parts, because this is what they are once the inverse Fourier transform is performed.

Of course, a similar phenomenon also exists in flat space. However, in flat space due to translation invariance, the action does not contain space-time coordinates explicitly, which makes integration by parts easy. Similarly, at the level of amplitudes, translation invariance entails the presence of the momentum-conserving delta functions appearing as an overall factor. This, in turn, leads to simple equivalence relations on the flat-space Mandelstam variables, which can be easily taken into account.

 In contrast, AdS space is not translationally invariant: the metric (\ref{MT}) and the Christoffel symbols (\ref{23oct20}) manifestly depend on space-time coordinates bringing this dependence into the action. 
In terms of amplitudes, this translates into the fact that the momentum-conserving delta functions no longer appear in the bare form, instead, they are supplemented with differential operators acting on them. Therefore, in AdS space momentum conservation does not hold  identically, rather, it holds up to terms that result from commuting the total momentum with the aforementioned differential operators. As a result, integration by parts becomes far less trivial, especially, when the number of derivatives in the vertex grows.
In what follows, we will not attempt to classify all possible forms of amplitudes, as this seems to be a tedious task. Instead, we will explore various forms of any given amplitude and try to find the most convenient one.

One thing that one can learn from example (\ref{20mar3}) is that derivatives of the delta function, that are contracted with momenta of fields on external lines can be eliminated. Such terms were absent for three-point amplitudes \cite{Nagaraj:2018nxq,Nagaraj:2019zmk} and it seems natural to try to achieve their absence for four-point amplitudes as well.
Another feature inherent to three-point amplitudes is that derivatives of the momentum-conserving delta function always appear in combination $\tilde G$. It is not hard to see that the same can be accomplished for (\ref{amp1}). Indeed, considering (\ref{20mar3}), the only term that has not yet been brought to such a form is the first one, for which we have
\begin{equation} \label{m3}
 -  \frac{\Box_p}{4R^4}\delta^{(4)}(p)=- \frac{p^2}{6}\tilde G^2\delta^{(4)}(p)+ \frac{2}{3R^2}\delta^{(4)}(p).
\end{equation}

Putting everything together, for the four-point amplitude (\ref{amp1}), we obtain
\begin{equation}
\label{20mar4}
    \begin{split}
       (2\pi)^{-4} A_4&=\left[-\frac{p^2}{6}+\frac{1}{2}p(p_1+p_2)-p_1p_2\right]\tilde G^2\delta^{(4)}(p)+\frac{2}{3R^2}\delta^{(4)}(p)\\
        &=\left[-\frac{1}{6}(s_1+s_2)+\frac{1}{12}(t_1+t_2)+\frac{1}{12}(u_1+u_2)\right]\tilde G^2\delta^{(4)}(p)+\frac{2}{3R^2}\delta^{(4)}(p),
    \end{split}
\end{equation}
where
\begin{equation}
\label{mandvar}
    \begin{split}
        &s_1\equiv 2p_1p_2=-\langle 12\rangle [12],\qquad s_2\equiv 2p_3p_4=-\langle 34\rangle [34],\\
        &t_1\equiv 2p_1p_4=-\langle 14\rangle [14],\qquad\, t_2\equiv 2p_2p_3=-\langle 23\rangle [23],\\
        &u_1\equiv 2p_1p_3=-\langle 13\rangle [13],\qquad u_2\equiv 2p_2p_4=-\langle 24\rangle [24]\\
    \end{split}
\end{equation}
are the AdS counterparts of the Mandelstam variables. Let us stress again that unlike in flat space, in AdS one cannot use
\begin{equation}
s_1= s_2, \qquad s_1+t_1+u_1= 0
\end{equation}
and other familiar identities. Indeed, e.g. $s_1-s_2$ vanishes only when multiplies the momentum-conserving delta function. Instead, in (\ref{20mar4}) the delta function is acted upon by $\tilde G^2$, so setting $s_1-s_2$ to zero is not legitimate.

One can note that (\ref{20mar4}) has one more property: the power of the $\tilde G$ operator acting on the delta function is twice the degree of the polynomial in the Mandelstam variables that appears in the prefactor. We will add this as an extra requirement for the sought form of the amplitude, which will be called canonical.

To summarize, we will say that the amplitude is written in the \emph{canonical form}, if it is presented as
\begin{equation}
\label{20mar5}
A_4 =  \sum_N f_N(s_1,s_2,t_1,t_2,u_1,u_2)\tilde G^{2N}\delta^{(4)}(p),
\end{equation}
where $f_N$ are  polynomials of the total homogeneity degree $N$ in the Mandelstam variables,
\begin{equation}
\label{20mar6}
\left(\frac{\partial}{\partial s_1}+\frac{\partial}{\partial s_2}+\frac{\partial}{\partial t_1}+\frac{\partial}{\partial t_2}+\frac{\partial}{\partial u_1}+\frac{\partial}{\partial u_2} \right)f_N = N f_N.
\end{equation}

To finish the discussion of two-derivative vertices, we note that all of them either reduce to (\ref{20mar7}) by permutations of fields' labels or to (\ref{vertex1}) by virtue of the free equations of motion. Moreover, one can see that by subtracting from (\ref{20mar4}) the amplitude for no-derivative interaction (\ref{trivial}) with a proper prefactor, one can eliminate the last term in (\ref{20mar4}). In other words, the $\tilde G^2$ part of (\ref{20mar4}) is a consistent amplitude on its own.

Having established the canonical form for two-derivative vertices, let us verify whether this can be achieved in more complicated cases. Consider a four-point vertex with four derivatives
\begin{equation}
\label{21mar2}
    S_4=\int d^4x\sqrt{-g}\nabla^{\mu}\phi_1\nabla_{\mu}\phi_2\nabla^{\nu}\phi_3\nabla_{\nu}\phi_4.
\end{equation}
Plugging in the metric and the scalar plane waves we get
\begin{equation} \label{1}
    \begin{split}
        S_4=\int d^4x &\Big(\frac{x^4}{16R^8}-\frac{ix^2}{8R^6}px G-\frac{x^2}{4R^4}(p_1p_2+p_3p_4)G^2\\
        &-\frac{1}{4R^4}(p_1+p_2)^i x_i(p_3+p_4)^jx_jG^2\\
        &+\frac{i}{2R^2}[p_1p_2(p_3+p_4)^ix_i+p_3p_4(p_1+p_2)^ix_i]G^3+p_1^ip_{2|i}p_3^jp_{4|j}G^4\Big)e^{ipx}.
    \end{split}
\end{equation}
Evaluating the Fourier transform and using
\begin{equation}
\label{21mar1}
\begin{split}
\frac{\partial}{\partial p^i}\tilde G^3\delta^{(4)}(p)&=-\frac{R^2}{2}p_i \tilde G^4\delta^{(4)}(p),\\
\frac{\partial}{\partial p^i}\frac{\partial}{\partial p^j}\tilde G^2\delta^{(4)}(p)&=\left(
\frac{R^4}{3}p_i p_j \tilde G^4 -\frac{2R^2}{5}\eta_{ij}\tilde G^2 - \frac{R^4}{30}\eta_{ij}p^2\tilde G^4\right)\delta^{(4)}(p),\\
\frac{\partial}{\partial p^i}\Box_p \tilde G\delta^{(4)}(p)&=
\left(\frac{12R^4}{5}p_i\tilde G^2-\frac{2R^6}{15}p_i p^2 \tilde G^4\right)\delta^{(4)}(p),\\
\Box_p^2\delta^{(4)}(p)&=\left(8R^4 - \frac{16 R^6}{5}p^2 \tilde G^2+\frac{R^8}{15}\tilde G^4 \right)\delta^{(4)}(p),
\end{split}
\end{equation}
we obtain
\begin{equation}
\label{21mar3}
  (2\pi)^{-4}  A_4=\frac{1}{2R^4}\delta^{(4)}(p)+\frac{1}{R^2}f_1\tilde G^2\delta^{(4)}(p)+f_2\tilde G^4\delta^{(4)}(p),
\end{equation}
where
\begin{equation}
    f_1=-\frac{1}{10}(s_1+s_2)+\frac{1}{20}(t_1+t_2)+\frac{1}{20}(u_1+u_2)
\end{equation}
and
\begin{equation}
    \begin{split}
        f_2=\frac{1}{16}\Big(&\frac{1}{5}s_1^2+\frac{26}{15}s_1s_2-\frac{2}{5}s_1t_1-\frac{2}{5}s_1t_2-\frac{2}{5}s_1u_1-\frac{2}{5}s_1u_2+\frac{1}{5}s_2^2\\
        &-\frac{2}{5}s_2t_1-\frac{2}{5}s_2t_2-\frac{2}{5}s_2u_1-\frac{2}{5}s_2u_2
        +\frac{1}{15}t_1^2+\frac{2}{15}t_1t_2+\frac{2}{15}t_1u_1\\
        &+\frac{2}{15}t_1u_2+\frac{1}{15}t_2^2+\frac{2}{15}t_2u_1+\frac{2}{15}t_2u_2
        +\frac{1}{15}u_1^2+\frac{2}{15}u_1u_2+\frac{1}{15}u_2^2\Big).
    \end{split}
\end{equation}
One can see that amplitude (\ref{21mar3}) is, indeed, in the canonical form.

In total,  there are three independent four-derivative vertices on-free-shell. All these can be given as linear combinations of (\ref{21mar2}) and two other vertices, generated by permutations of field's labels. Hence, we conclude that all four-derivative amplitudes in AdS space admit the canonical form. Moreover, as in the case of the two-derivative amplitude, it is straightforward to see that the $\tilde G^4$ part of (\ref{21mar3}) is consistent on its own.

As a last example, we consider a six-derivative vertex
\begin{equation}
\label{21mar4}
    S_4=\int d^4x\sqrt{-g}\nabla^\rho\nabla^{\mu}\phi_1\nabla_{\mu}\phi_2\nabla_\rho\nabla^{\nu}\phi_3\nabla_{\nu}\phi_4.
\end{equation}
It turns out, that the associated amplitude can be brought to the canonical form as well.
The result of this computation is lengthy so we give it in  appendix \ref{appa:6der}. Again, by permuting fields' labels one obtains other amplitudes, which together generate all independent amplitudes with six derivatives. Moreover, as in previous examples,  the piece of the amplitude for (\ref{21mar4}) with the highest power of $\tilde G$ -- which is six in this case -- is a consistent amplitude by itself.

To summarize, in this section we evaluated and explored various forms of lower-derivative four-point amplitudes in AdS${}_4$ in the spinor-helicity representation. We found that in all cases we considered, amplitudes can be brought to the canonical form, which is defined in  (\ref{20mar5}), (\ref{20mar6}). Moreover, our experience shows, that for every amplitude, its canonical form is unique. In the following, we will assume that these two properties of the canonical form -- that it can be always achieved for consistent amplitudes and that it is unique -- are true in general. The proofs of these two properties statements look feasible, but technical and we leave them for future research.

Finally, we would like to stress again that all forms of amplitudes are equally valid. We chose to deal with the canonical form, because it is simple and allows to fix the freedom of integration by parts. This, in turn, is helpful for developing systematics for the AdS four-point amplitudes in the spinor-helicity representation, such as their classification, which will be performed in the next section.

\section{Amplitudes from Symmetries} \label{AFS}
In the previous section, we used scalar plane waves to calculate certain simple 4-particle amplitudes. In this section, we will use symmetry arguments to construct all contact four-point amplitudes of scalar fields. Analogous arguments were used in 
\cite{Nagaraj:2019zmk} to classify three-point amplitudes of spinning fields.

Lorentz invariance implies that all Lorentz indices should be contracted covariantly. Hence, our ansatz for the four-point amplitude should involve arbitrary functions of all Lorentz scalars that one can construct out of momenta of fields involved and their derivatives.
However, in the previous section we saw that dependence on some of these variables can be removed using integration by parts. Moreover, it is of more interest to classify inequivalent amplitudes, rather than all amplitudes, which include inequivalent amplitudes and all their forms that can be achieved using integration by parts. Therefore, below we will take into account the intuition gained in the previous section and study a more constrained ansatz. Our goal will be to show that amplitudes in the canonical form transform properly under AdS isometries, provided certain constraints are imposed on the polynomials of the Mandelstam variables, that this form features. Then, we will solve these constraints and show that there are as many solutions to these constraints as vertices in AdS, that are independent on-shell.

So, we consider an asatz
\begin{equation}
\label{23mar1}
    {A} = \sum_N f_N(s_1,s_2,t_1,t_2,u_1,u_2)g_N(\Box_p)\delta^{(4)}(p).
\end{equation}
With the Lorentz invariance taken into account, it remains to impose invariance with respect to deformed translations,
\begin{equation} \label{SymmetryEq}
    \left(\mathcal{P}_{1|\alpha\dot{\alpha}}+\mathcal{P}_{2|\alpha\dot{\alpha}}+\mathcal{P}_{3|\alpha\dot{\alpha}}+\mathcal{P}_{4|\alpha\dot{\alpha}}\right){A}=0,
\end{equation}
which were defined in (\ref{isometry}). Direct evaluation gives
\begin{equation}
\label{23mar2}
    \begin{split}
        &\sum_N \Big[f_N\left(p^c+\frac{2}{R^2}\frac{\partial}{\partial p_c}+\frac{1}{2R^2}p^a\frac{\partial}{\partial p^a}\frac{\partial}{\partial p_c}-\frac{1}{4R^2}p^c\Box_p\right)g_N\left(\Box_p\right)\delta^{(4)}(p)\\
        &\quad+\frac{1}{R^2}\left(s_1\frac{\partial }{\partial s_1}+s_2\frac{\partial }{\partial s_2}+t_1\frac{\partial }{\partial t_1}+t_2\frac{\partial }{\partial t_2}+u_1\frac{\partial }{\partial u_1}+u_2\frac{\partial }{\partial u_2}\right)f_N\frac{\partial}{\partial p_c}g_N(\Box_p)\delta^{(4)}(p)\\
        &\quad\qquad\qquad\qquad\qquad\qquad\qquad\qquad\quad+\frac{1}{2R^2}\Bar{\sigma}^{c\dot{\alpha}\alpha}\sum_{m=1}^4\frac{\partial^2 f_N}{\partial\lambda_m^{\alpha}\partial \Bar{\lambda}_m^{\dot{\alpha}}}g_N(\Box_p)\delta^{(4)}(p)\Big]=0.
    \end{split}
\end{equation}
Taking into account (\ref{20mar6}), the first two lines in (\ref{23mar2}) give
\begin{equation}
\label{23mar3}
\begin{split}
 &\sum_N f_N\left(p^c+\frac{N+2}{R^2}\frac{\partial}{\partial p_c}+\frac{1}{2R^2}p^a\frac{\partial}{\partial p^a}\frac{\partial}{\partial p_c}-\frac{1}{4R^2}p^c\Box_p\right)g_N\left(\Box_p\right)\delta^{(4)}(p)\\
&\quad\qquad\qquad\qquad\qquad\quad =  \sum_N f_N\left(\left(\Box_p+4R^2\right)g_N'\left(\Box_p\right)-2Ng_N\left(\Box_p\right)\right)\frac{\partial \delta^{(4)}(p)}{\partial p_c}.
\end{split}
\end{equation}
For 
\begin{equation}
\label{23mar4}
g_N(\Box_p)=\tilde G^{2N},
\end{equation}
which is its value in the canonical form (\ref{20mar5}), the bracket in the last line of (\ref{23mar3}) vanishes. 
So the AdS covariance condition (\ref{23mar2}) reduces to
\begin{equation} \label{constraints}
  \sum_N  \sum_{m=1}^4\frac{\partial^2 f_N}{\partial\lambda_m^{\alpha}\partial \Bar{\lambda}_m^{\dot{\alpha}}}g_N(\Box_p)\delta^{(4)}(p)=0
\end{equation}
with $g_N$ given in (\ref{23mar4}).

Equation (\ref{constraints}) has four components, which are labelled by $\alpha$ and $\dot\alpha$. To extract these components, we contract (\ref{constraints}) with four momenta $\lambda_n^\alpha\bar\lambda^{\dot\alpha}_n$, $n=1,2,3,4$. Next, we use integration by parts and the spinor algebra to get rid off all spinor products that cannot be expressed in terms of the Mandelstam variables. This computation is rather straightforward, but tedious, so we present it in appendix \ref{app3:constraints}. Eventually, we find that the AdS-covariance conditions imply
\begin{equation}
\label{23mar5}
\begin{split}
{\cal D}(s_1,s_2,t_1,t_2,u_1,u_2)f_N=0,\\
{\cal D}(s_1,s_2,u_2,u_1,t_2,t_1)f_N=0,\\
{\cal D}(t_2,t_1,s_2,s_1,u_1,u_2)f_N=0,\\
{\cal D}(u_2,u_1,t_1,t_2,s_2,s_1)f_N=0,
\end{split}
\end{equation}
where the operator ${\cal D}$ is given by
\begin{equation}
\label{23mar6}
\begin{split}
{\cal D}(s_1,s_2,t_1,t_2,u_1,u_2)&\equiv
\left(s_1\frac{\partial}{\partial s_1}+t_1\frac{\partial}{\partial t_1}+u_1\frac{\partial}{\partial u_1}\right)^2+(t_1+u_1)\frac{\partial}{\partial s_2}s_2\frac{\partial}{\partial s_2}\\
&+(s_1+u_1)\frac{\partial}{\partial t_2}t_2\frac{\partial}{\partial t_2}+(s_1+t_1)\frac{\partial}{\partial u_2}u_2\frac{\partial}{\partial u_2}\\
&+(t_1t_2+u_1u_2-s_1s_2)\frac{\partial^2 }{\partial t_2 \partial u_2}+
(s_1s_2+t_1t_2-u_1u_2)\frac{\partial^2 }{\partial t_2 \partial s_2}\\
&+(s_1s_2+u_1u_2-t_1t_2)\frac{\partial^2 }{\partial s_2 \partial u_2}.
\end{split}
\end{equation}

Before proceeding with the solution of (\ref{23mar5}), let us note that operators ${\cal D}$ commute with the total homogeneity degree in the Mandelstam variables, which means that solutions with different $N$ are independent. This is consistent with our observations from the explicit amplitude computations in the previous section. One can also check that all amplitudes we computed so far do satisfy (\ref{23mar5}). From now on, we will drop label $N$ for notational simplicity.

In flat space, there are $N+1$ independent amplitudes of the homogeneity degree $N$ in the Mandelstam variables. These can be given, for example, as polynomials in $s$ and $t$, with the $u$ dependence eliminated by momentum conservation. Similarly, in the AdS case, when the total homogeneity degree in the Mandelstam variables is fixed to $N$, we should find $N+1$ independent amplitudes. One can label these in different ways. To strengthen the analogy with the flat-space case, we will label the solutions to (\ref{23mar5}) by their values at a hypersurface where four out of six Mandelstam variables are set to zero, that is by a polynomial of the two remaining Mandelstam variables. For example, we can set the boundary data as
\begin{equation}
\label{23mar7}
f(s_1,0,t_1,0,0,0)=h(s_1,t_1).
\end{equation}
Below we will show that (\ref{23mar5}) allow to reconstruct $f$ completely once the boundary condition (\ref{23mar7}) is set.

To do that, let us first consider the third equation in (\ref{23mar5}). More specifically, we will be interested only in the equations, that are $s_2$-, $t_2$- and $u_2$-independent. In other words, we focus on a sector of
\begin{equation}
\label{23mar8}
{\cal D}(t_2,t_1,s_2,s_1,u_1,u_2)f=0
\end{equation}
that satisfies
\begin{equation}
\label{23mar9}
\hat N_{s_2}{\cal D}f=0, \qquad \hat N_{t_2}{\cal D}f=0, \qquad \hat N_{u_2}{\cal D}f=0.
\end{equation}
Here we use the notation 
\begin{equation}
\label{23mar10}
\begin{split}
\hat N_{s_2}\equiv s_2\frac{\partial}{\partial s_2}, \qquad 
\hat N_{t_2}\equiv t_2\frac{\partial}{\partial t_2}, \qquad 
\hat N_{u_2}\equiv u_2\frac{\partial}{\partial u_2},
 \\
\hat N_{s_1}\equiv s_1\frac{\partial}{\partial s_1}, \qquad 
\hat N_{t_1}\equiv t_1\frac{\partial}{\partial t_1}, \qquad 
\hat N_{u_1}\equiv u_1\frac{\partial}{\partial u_1}.
\end{split}
\end{equation}
If we take into account that $f$ is a polynomial, it is not hard to see that most of terms in (\ref{23mar8}) drop out. Consider, for example,
\begin{equation}
\label{23mar11}
s_1s_2 \frac{\partial^2}{\partial s_1\partial u_2}f \quad \subset \quad  {\cal D}f.
\end{equation}
Since, we demand that ${\cal D}f$ has the homogeneity degree zero in $s_2$, for $f$ to contribute to it via the term on the left hand side of (\ref{23mar11}), the homogeneity degree of $f$ in $s_2$ should be minus one. However,  this is impossible, since $f$ is a polynomial. By using the same type of arguments, we find that the only terms that remain, lead to
\begin{equation}
\label{23mar12}
\left(\hat N_{u_1}^2+u_1\frac{\partial}{\partial t_1}t_1\frac{\partial}{\partial t_1}+u_1\frac{\partial}{\partial s_1}s_1\frac{\partial}{\partial s_1}\right)f(s_1,0,t_1,0,u_1,0)=0.
\end{equation}

Next, it is not hard to see that 
\begin{equation}
\label{23mar13}
[\hat N_{u_1},\hat O_1] = \hat O_1,
\end{equation}
where we denoted
\begin{equation}
\label{23mar13x1}
\hat O_1\equiv -u_1\frac{\partial}{\partial t_1}t_1\frac{\partial}{\partial t_1}-u_1\frac{\partial}{\partial s_1}s_1\frac{\partial}{\partial s_1}.
\end{equation}
Considering that 
\begin{equation}
\label{23mar14}
\hat N_{u_1}f(s_1,0,t_1,0,0,0)=0,
\end{equation}
we can solve for the $u_1$-dependence of $f$ from the boundary data (\ref{23mar7})
by expanding it in powers of $u_1$ and solving (\ref{23mar12}) order by order.
 As a result, one finds
\begin{equation}
\label{23mar15}
f(s_1,0,t_1,0,u_1,0) = I_0(2\sqrt{\hat O_1})h(s_1,t_1)\equiv \sum_{n=0}^\infty \frac{\hat O_1^n}{(n!)^2}h(s_1,t_1).
\end{equation}
Though, this formula features an infinite sum, it, actually, truncates to the first $N+1$ terms due to the fact that $h$ is a polynomial of $N$th degree.

It is hard to proceed further directly. We can note, however, that all amplitudes that we computed in the previous section enjoy symmetry with respect to three independent permutations of the Mandelstam variables $s_1\leftrightarrow s_2$, $t_1\leftrightarrow t_2$,  and $u_1\leftrightarrow u_2$. It seems reasonable to expect that this symmetry holds in general.
Assuming that this symmetry does hold,  we consider the first equation in (\ref{23mar5}) with $\{s_1,t_1,u_1 \}$ and $\{s_2,t_2,u_2 \}$ interchanged. This gives
\begin{equation}
\label{23mar16}
\left(\hat N_{s_2}+\hat N_{t_2}+\hat N_{u_2}\right)^2f- \hat O_2f=0,
\end{equation}
where
\begin{equation}
\label{23mar17}
\begin{split}
\hat O_2\equiv &-(t_2+u_2)\frac{\partial}{\partial s_1}s_1\frac{\partial}{\partial s_1}-(s_2+u_2)\frac{\partial}{\partial t_1}t_1\frac{\partial}{\partial t_1}-(s_2+t_2)\frac{\partial}{\partial u_1}u_1\frac{\partial}{\partial u_1}\\
&-(-s_1s_2+t_1t_2+u_1u_2)\frac{\partial^2 }{\partial t_1 \partial u_1}-
(s_1s_2+t_1t_2-u_1u_2)\frac{\partial^2 }{\partial t_1 \partial s_1}\\
&-(s_1s_2-t_1t_2+u_1u_2)\frac{\partial^2 }{\partial s_1 \partial u_1}.
\end{split}
\end{equation}
Noting that 
\begin{equation}
\label{23mar18}
[\hat N_{s_2}+\hat N_{t_2}+\hat N_{u_2},\hat O_2]=\hat O_2
\end{equation}
and
\begin{equation}
\label{23mar19}
\left(\hat N_{s_2}+\hat N_{t_2}+\hat N_{u_2}\right) f(s_1,0,t_1,0,u_1,0)=0,
\end{equation}
we solve for $f$ as
\begin{equation}
\label{23mar20}
\begin{split}
f(s_1,s_2,t_1,t_2,u_1,u_2)&= I_0(2\sqrt{\hat O_2})f(s_1,0,t_1,0,u_1,0) \\
&= I_0(2\sqrt{\hat O_2}) I_0(2\sqrt{\hat O_1})h(s_1,t_1).
\end{split}
\end{equation}

The method of solving (\ref{23mar5}) that we presented here is just a compact way to summarize our findings obtained with the Frobenius method. Unfortunately, more explicit evaluation of the right hand side in (\ref{23mar20}) is obstructed by the fact that $\hat O_2$ is given by a sum of operators, that do not commute. Still, representation (\ref{23mar20}) can be used rather efficiently to generate solutions with low $N$ from the boundary data. 

It is worth stressing that, while solving equations, we only used some of them. We also used the symmetry of the solutions with respect to $\{s_1,t_1,u_1 \}\leftrightarrow \{s_2,t_2,u_2 \}$, which was not derived from the equations, but observed from particular solutions. This means that one still has to verify that (\ref{23mar20}) does solve (\ref{23mar5}). It does not seem to be easy to do that in general, though, we checked this for particular examples.

\section{Derivation from Conformal Primaries}
\label{sec5:weyl}

In the previous sections we saw that AdS amplitudes after certain manipulations have the same form as flat-space amplitudes multiplied with the appropriate  powers of the AdS conformal factor. It is natural to try to connect this property of amplitudes with the behaviour of the associated vertices under Weyl transformations. Due to our choice of AdS plane waves, Weyl-invariant vertices have identical amplitudes in flat and AdS spaces. Even if the vertex is not Weyl-invariant, but scales with a certain power of the conformal factor, it can be made Weyl-invariant by multiplying it with an auxiliary field, that scales appropriately to compensate scaling of the vertex.
As a result, the associated amplitudes in AdS and flat spaces are equal up to a power of the AdS conformal factor. Moreover, for this relation to hold, we do not need to require that vertices transform appropriately under general Weyl transformations, instead,  it suffices to know how they behave under the Weyl transformation that maps flat space to AdS.

To find out how vertices transform with respect to this particular Weyl transformation and select the appropriate ones, we would need to do a separate analysis. Instead, we will use that Weyl invariance is typically connected to conformal invariance and, instead, consider vertices that transform as conformal primaries. In the latter case, many relevant results are already available. It is worth keeping in mind that conformal invariance does not necessarily imply Weyl invariance -- some counterexamples can be found, e.g. in \cite{GJMS}. Therefore, without doing further analysis it is not guaranteed that our shortcut can be used in general, though, as we found, it does allow to produce consistent AdS amplitudes in the examples that we considered.

More explicitly, let us assume that we are given a Lagrangian density 
\begin{equation}
\label{27mar1}
L(\partial,\eta,\phi_m)=L(\partial_1, \partial_2,\partial_3,\partial_4,\phi_1,\phi_2,\phi_3,\phi_4),
\end{equation}
which is linear in each of the four massless fields $\phi_i$ and, in addition, under conformal transformation it transforms as a scalar conformal primary of dimension $\Delta=2N+4$, where $2N$ is the number of derivatives $L$ features. Then, by employing an auxiliary massless field $\phi_0$, we can construct a conformally invariant vertex
\begin{equation}
\label{27mar2}
S[\partial,\eta,\phi_m,\phi_0]=\int d^4 x L(\partial,\eta,\phi_m)\, \phi_0^{-2N}.
\end{equation}
By our assumption, it can be promoted to a vertex of the form
\begin{equation}
\label{27mar3}
S'[\nabla,g,\phi_m,\phi_0]=\int d^4 x\sqrt{-g} L(\nabla,g,\phi_m)\, \phi_0^{-2N}+\dots,
\end{equation}
which is invariant with respect to the Weyl transformation, that maps flat space to AdS. Here $\dots$ refer to terms, that involve the curvature tensor.

Let us now consider a flat-space amplitude in background $\phi_0=1$
\begin{equation}
\label{27mar5}
A\equiv S[\partial,\eta,e^{ip_mx},1].
\end{equation} 
 This amplitude takes the form
\begin{equation}
\label{27mar6}
A=(2\pi)^4f(s_1,s_2,t_1,t_2,u_1,u_2)\delta^{(4)}(p).
\end{equation}
Note that when evaluating (\ref{27mar6}) we should not use momentum conservation or, equivalently, integration by parts. The reason is that we would like to avoid derivatives acting on $\phi_0$, which is necessary for our argument to work.

Then, we make the Weyl transformation that relates flat and AdS space metrics. Weyl invariance of the vertex (\ref{27mar3}) implies that in AdS space the amplitude remains the same
\begin{equation}
\label{27mar7}
A'\equiv S'[\nabla_{\rm AdS},g_{\rm AdS},G e^{ip_mx},G]=S[\partial,\eta,e^{ip_mx},1]=A.
\end{equation}

Using this result, it is straightforward to compute a usual four-point amplitude in AdS
\begin{equation}
\label{27mar8}
A''\equiv \int d^4 x\sqrt{-g_{\rm AdS}} L(\nabla_{\rm AdS},g_{\rm AdS},Ge^{ip_mx})+\dots,
\end{equation}
which is of the form, we considered in section \ref{AUPW}.
Namely, this amplitude differs from $A'$ only  by an overall factor, contributed by  $\phi_0$. Taking this difference into account, we obtain
\begin{equation}
\label{28mar9}
A''=S'[\nabla_{\rm AdS},g_{\rm AdS},G e^{ip_mx},1] =(2\pi)^4f(s_1,s_2,t_1,t_2,u_1,u_2)\tilde G^{2N}\delta^{(4)}(p).
\end{equation}

In other words, we find that there is a class of contact four-point vertices in AdS space that gives amplitudes immediately in the canonical form. Moreover, for these vertices, one can avoid a tedious AdS computation: by employing Weyl transformation the computation reduces to the flat-space one, which is much simpler.

To illustrate this idea, let us consider 
\begin{equation}
\label{28mar10}
L(\partial,\eta,\phi_m)=J_{ab}(\phi_1,\phi_2)J^{ab}(\phi_3,\phi_4),
\end{equation}
where $J$ is the traceless and symmetric spin-$2$ conserved current
\begin{equation}
\label{28mar11}
J_{ab}(\phi_m,\phi_n) =2(\partial_a\partial_b\phi_m \phi_n+\phi_m \partial_a\partial_b\phi_n)-
4(\partial_a\phi_m\partial_b\phi_n+\partial_a\phi_n\partial_b\phi_m)+\eta_{ab}\partial^c\phi_m\partial_c\phi_n.
\end{equation}
By a straightforward evaluation, we find that $A$ defined in (\ref{27mar5}) equals (\ref{27mar6}) with 
\begin{equation}
\label{28mar12}
f=-4s_1s_2+t_1^2+8t_1t_2+t_2^2-4t_1u_1-4t_2u_1+u_1^2-4t_1u_2-4t_2u_2+8 u_1u_2+u_2^2.
\end{equation}
Then, taking into account the contribution from the conformal factor, we obtain the AdS amplitude
\begin{equation}
\label{28mar13}
A''=(2\pi)^4f(s_1,s_2,t_1,t_2,u_1,u_2)\tilde G^{4}\delta^{(4)}(p).
\end{equation}
One can check that this result is consistent with the symmetry constraints from section \ref{AFS}: it has the correct power of the conformal factor and $f$ satisfies (\ref{23mar5}).

In an analogous manner we considered a Lagrangian with two spin-1 currents contracted as well as  Lagrangians of the form\footnote{Explicit expressions for primaries (\ref{17apr1}) can be found in \cite{Bekaert:2015tva}.}
\begin{equation}
\label{17apr1}
\Box^n (\phi_1\phi_2) \phi_3\phi_4+\dots,
\end{equation}
up to $n=10$ and found that the resulting amplitudes, indeed, solve the constraints from the previous section.

Finally, we note that the set of conformal primaries (\ref{27mar1}) is large enough to provide a basis for amplitudes in flat space and, hence, in AdS space as well. Indeed, one can consider primaries of the form
\begin{equation}
\label{28mar14}
L_{n,l} = \Box^n J_{a_1\dots a_l}(\phi_1,\phi_2) J^{a_1\dots a_l}(\phi_3,\phi_4)+\dots, 
\end{equation}
where $J$'s are conserved, traceless and symmetric spin-$l$ currents and $\dots$ refer to other terms with $2N$ derivatives, that are necessary to make $L$ primary. By construction, the associated flat-space amplitude has spin $l$ in the $s$ channel. Moreover, no matter what is the exact way the derivatives are distributed in the terms that we omitted in (\ref{28mar14}), they contribute an extra factor of $s^n$ to the amplitude. In other words,  four-point vertex (\ref{28mar14}) results in a flat-space amplitude of the form
\begin{equation}
\label{28mar15}
A\propto  s^{n+l}P_l(\cos \theta)\delta^{(4)}(p), \qquad \cos \theta \equiv \frac{t-u}{s},
\end{equation}
where $P_l$ are the Legendre polynomials. It is well-known that (\ref{28mar15}) provides a basis in the space of polynomial flat-space amplitudes. For example, keeping $n+l=N$ fixed and changing $l$ from $0$ to $N$, we obtain a basis of $N+1$ elements in the space of polynomial amplitudes of order $N$ in the Mandelstam variables.

\section{Conclusion and Outlook}
\label{sec6:conc}

In the present paper we explored AdS${}_4$ spinor-helicity amplitudes for contact four-point diagrams from different angles. To start, we computed a number of lower-derivative amplitudes employing the standard Feynman rules. At a technical level, this involves evaluation of some simple Fourier transforms. The key difference of this computation with its flat space counterpart is that due to the absence of translation invariance in AdS space, the action manifestly depends on space-time coordinates, and, as a result, amplitudes involve derivatives of the momentum-conserving delta-function. This, in turn, implies that momentum is not conserved in AdS space and the associated machinery has to be deformed. In particular, one cannot simply trade the Mandelstam variables one for another in the standard manner to bring amplitudes to a more convenient form. In AdS, similar equivalence relations between different forms of amplitudes still exist, but become  more complicated.

In the first part of the paper we explored various forms of amplitudes and found that each amplitude can be brought to the form, which is especially simple. We call this form canonical. In this form, all derivatives of the momentum-conserving delta-function combine into powers of the Fourier transform of the AdS conformal factor, while the remaining part of the amplitude is given by a polynomial in the Mandelstam variables -- just like in flat space.
 We then used the canonical form to classify consistent amplitudes associated with contact four-point diagrams, by requiring correct transformation properties with respect to the AdS isometry algebra. The result of this analysis is consistent with our expectations: in particular, we find that, as in flat space, contact four-point amplitudes can be labelled by polynomials of two variables. Finally, we establish a connection between the canonical form of amplitudes and scalar conformal primary operators constructed out of four massless fields. 

Our main motivation in this paper was to make the first step towards understanding the relation between the analytic structure of AdS amplitudes in the spinor-helicity representation and the type of the diagram they result from. In this respect, we can conclude that amplitudes for contact diagrams do have distinctive analytic structure: leaving aside the AdS conformal factor, they are given by polynomials of the Mandelstam variables. An obvious next step would be to compute amplitudes for exchange diagrams and compare their analytic structure with what we found here for contact interactions. Eventually, these results may serve as a basis for the development of the on-shell methods for the AdS spinor-helicity representation in future. It would also be interesting to explore AdS generalizations of other modern amplitude techniques such as the color-kinematics duality \cite{Bern:2008qj}\footnote{See \cite{Li:2018wkt,Farrow:2018yni} for discussions of the color-kinematics duality in the context of the AdS/CFT correspondence.} and the CHY formalism \cite{Cachazo:2013gna,Cachazo:2013hca}. 

Finally, let us note that this paper was devoted to four-point amplitudes of scalar fields, while the power of the spinor-helicity formalism becomes more apparent when one deals with fields with spin. Though, we expect that, as in flat space, spinning contact diagrams have similar analytic structure to scalar ones, it would be interesting to consider them in future.

\acknowledgments

We would like to thank E. Skvortsov for useful comments on the draft. The work of B. N. was supported by the Mitchell/Heep Chair in High Energy Physics. The work of D. P. was supported in part by  RSF Grant 18-12-00507.

\appendix

\section{Conventions and Useful Formulae}
\label{appa:conventions}

In this appendix we collect our notations and some formulae used in the text. For more details we refer the reader to \cite{Nagaraj:2019zmk}.

We are dealing with the four-dimensional Minkowski and AdS spaces in the mostly plus signature. For vector Lorentz indices we use letters from the beginning of the Latin alphabet, while letters from the middle of the Latin alphabet are used to label particles. We also use Greek letters from the beginning of the alphabet for spinor indices and Greek letters from the middle of the alphabet for base indices. 

We use the following conventions for the Pauli matrices
\begin{equation}
\label{5nov3}
\sigma^0 = 
\left(\begin{array}{cccc}
1 &&& 0\\
0 && &1
\end{array}\right), \quad \sigma^1 = 
\left(\begin{array}{cccc}
0 & && 1\\
1& && 0
\end{array}\right), 
\quad 
 \sigma^2 = 
\left(\begin{array}{ccc}
0 && -i\\
i&& 0
\end{array}\right), \quad
 \sigma^3 = 
\left(\begin{array}{ccc}
1 && 0\\
0&& -1
\end{array}\right).
\end{equation}
These allow to convert Lorentz vector indices to spinor ones and vice versa. For example,
\begin{equation}
\label{5nov4}
p_{\alpha\dot\alpha}\equiv p_a (\sigma^a)_{\alpha\dot\alpha}, \qquad  p_a=-\frac{1}{2}(\sigma_a)^{\dot\alpha\alpha}p_{\alpha\dot\alpha}.
\end{equation}
We raise and lower spinor indices according to 
\begin{equation}
\label{5nov6}
\lambda^\alpha = \epsilon^{\alpha\beta} \lambda_\beta, \qquad \lambda_\beta=\epsilon_{\beta\gamma} \lambda^\gamma,
\end{equation}
where
\begin{equation}
\label{5nov7}
\epsilon^{\alpha\beta}=\epsilon^{\dot\alpha\dot\beta}=
\left(
\begin{array}{cccc}
0&&& 1\\
-1&&&0
\end{array}
\right) = -\epsilon_{\alpha\beta}=-\epsilon_{\dot\alpha\dot\beta}.
\end{equation}
Derivatives with respect to spinors are defined in a natural way
\begin{equation}
\label{5nov15}
\frac{\partial \lambda^\alpha}{\partial \lambda^\beta}= \delta^\alpha_\beta, \qquad
\frac{\partial \lambda_{\alpha}}{\partial \lambda_{\beta}}= \delta_{\alpha}^{\beta}.
\end{equation}

In our conventions one has
 \begin{equation}
 \label{5nov9}
 \begin{split}
 (\sigma^a)_{\alpha\dot\alpha}(\sigma_a)_{\beta\dot\beta}&=-2\epsilon_{\alpha\beta}\epsilon_{\dot\alpha\dot\beta},
 \qquad
  (\sigma^a)^{\alpha\dot\alpha}(\sigma_a)^{\beta\dot\beta}=-2\epsilon^{\alpha\beta}\epsilon^{\dot\alpha\dot\beta},\\
  (\sigma_{a})_{\alpha\dot{\alpha}}({\sigma}^{b})^{\dot{\alpha}\alpha}&=-2\delta^{b}_{a}, \qquad\qquad
   A_{\alpha\beta}-A_{\beta\alpha}=\epsilon_{\alpha\beta}A^\gamma{}_\gamma,\\
  ({\sigma}^a)^{\dot\alpha\beta}(\sigma^b)_{\beta\dot\beta}+({\sigma}^b)^{\dot\alpha\beta}(\sigma^a)_{\beta\dot\beta}&=-2\eta^{ab}\delta^{\dot{\alpha}}_{\dot{\beta}},\\
  (\sigma^a)_{\beta\dot\alpha}(\sigma^c)^{\dot\alpha\alpha}(\sigma^b)_{\alpha\dot\gamma}+(a\leftrightarrow b)&=
-2\left(\eta^{ac}(\sigma^b)_{\beta\dot\gamma}+\eta^{cb}(\sigma^a)_{\beta\dot\gamma}-\eta^{ab}(\sigma^c)_{\beta\dot\gamma}\right).
  \end{split}
 \end{equation}

In addition, we use the shorthand notations
\begin{equation}
\label{5nov19}
\begin{split}
\langle mn \rangle&\equiv \lambda^m_\alpha\lambda^{n\alpha}=\lambda^i_\alpha\lambda^j_\beta\epsilon^{\alpha\beta}, \qquad
\left[ mn \right]\equiv \bar{\lambda}^m_{\dot{\alpha}}\bar{\lambda}^{n\dot{\alpha}}=\bar{\lambda}^m_{\dot{\alpha}}\bar{\lambda}^n_{\dot{\beta}}\epsilon^{\dot{\alpha}\dot{\beta}},\\
\langle m x n] &\equiv \lambda_m^\alpha x_{\alpha\dot\alpha} \bar\lambda^{\dot\alpha}_n,
\qquad \qquad\quad \;\;\langle \lambda x \mu ] \equiv \lambda^\alpha x_{\alpha\dot\alpha}\bar\mu^{\dot\alpha}.
\end{split}
\end{equation}

\section{Computation of a Six-derivative Amplitude}
\label{appa:6der}

In this section we give some intermediate results and useful relation, relevant for the computation of the canonical form of the amplitude for vertex  (\ref{21mar4}). 

As usual, we start by plugging in the plane wave solutions and expressions for the metric and the Christoffel symbols into the vertex. Next, we evaluate the Fourier transform. Using that
\begin{equation}
\label{21mar5}
\begin{split}
\frac{\partial}{\partial p^i}\tilde G^5\delta^{(4)}(p)&=-\frac{R^2}{3}p_i \tilde G^6\delta^{(4)}(p),\\
\frac{\partial}{\partial p^i}\frac{\partial}{\partial p^j}\tilde G^4\delta^{(4)}(p)&=\left(
\frac{2R^4}{15}p_i p_j \tilde G^6 -\frac{2R^2}{7}\eta_{ij}\tilde G^4 - \frac{R^4}{105}\eta_{ij}p^2\tilde G^6\right)\delta^{(4)}(p),\\
\frac{\partial}{\partial p^i}\frac{\partial}{\partial p^j}\frac{\partial}{\partial p^k} \tilde G^3\delta^{(4)}(p)&=
\left(
\frac{R^4}{7}(\eta_{jk}p_i+\eta_{ik}p_j+\eta_{ij}p_k)\tilde G^4\right.
\\&\quad\left.-\frac{R^6}{15}p_ip_jp_k \tilde G^6
+\frac{R^6}{210}(\eta_{jk}p_i+\eta_{ik}p_j+\eta_{ij}p_k)p^2\tilde G^6
\right)\delta^{(4)}(p),\\
\frac{\partial}{\partial p^i}\frac{\partial}{\partial p^j}\Box_p\tilde G^2\delta^{(4)}(p)&=\left(
\frac{4R^4}{5}\eta_{ij}\tilde G^2 -\frac{16 R^6}{21}p_i p_j \tilde G^4 +\frac{2 R^8}{105}p_i p_j p^2 \tilde G^6\right.\\
&\qquad\qquad\left.
+\frac{2R^6}{105}\eta_{ij}p^2 \tilde G^4 -\frac{R^8}{630}\eta_{ij}p^4 \tilde G^6
 \right)\delta^{(4)}(p),\\
 \frac{\partial}{\partial p_i}\Box_p^2\tilde G \delta^{(4)}(p)&=
 \left(-\frac{32 R^6}{5}p_i \tilde G^2 +\frac{64 R^8}{105}p_i p^2 \tilde G^4-\frac{2R^{10}}{315}p_i p^4 \tilde G^6 \right)\delta^{(4)}(p),\\
 \Box_p^3\delta^{(4)}(p)&=\left(-\frac{128R^6}{5}+ \frac{64 R^8}{5}p^2 \tilde G^2-\frac{16 R^{10}}{35}p^4 \tilde G^4+\frac{4R^{12}}{1575}p^6 \tilde G^6 \right)\delta^{(4)}(p),
\end{split}
\end{equation}
we find the associated amplitude to be
\begin{equation}
\label{21mar6}
 (2\pi)^{-4}A_4=\frac{19}{480R^6}\delta^{(4)}(p)+\frac{1}{R^4}f_1\tilde G^2\delta^{(4)}(p)+\frac{1}{R^2}f_2\tilde G^4\delta^{(4)}(p)+
 f_3\tilde G^6\delta^{(4)}(p),
\end{equation}
where 
\begin{equation}
\label{21mar7}
\begin{split}
f_1&=-\frac{1}{240}(s_1+s_2-17 t_1-17 t_2+16 u_1 + 16 u_2),\\
f_2&=\frac{1}{6720}(57 s_1^2 + 494 s_1 s_2 + 57 s_2^2 - 202 s_1 t_1 - 202 s_2 t_1 + 41 t_1^2
 -  202 s_1 t_2- 202 s_2 t_2  \\ 
 &\qquad\qquad+ 302 t_1 t_2 + 41 t_2^2 - 26 s_1 u_1 - 26 s_2 u_1
 +  38 t_1 u_1 + 38 t_2 u_1- 3 u_1^2  \\
 &\qquad\qquad \qquad\qquad\qquad\;\;- 26 s_1 u_2   - 26 s_2 u_2 + 38 t_1 u_2 + 
 38 t_2 u_2 - 226 u_1 u_2 - 3 u_2^2),\\
 f_3&=\frac{1}{16800}(6 s_1^3 + 198 s_1^2 s_2 + 198 s_1 s_2^2 + 6 s_2^3 - 2 s_1^2 t_1 - 
 34 s_1 s_2 t_1 - 2 s_2^2 t_1 - 7 s_1 t_1^2\\
 &\qquad\qquad\quad- 7 s_2 t_1^2  + t_1^3 - 2 s_1^2 t_2 - 
 34 s_1 s_2 t_2 - 2 s_2^2 t_2 - 164 s_1 t_1 t_2 - 164 s_2 t_1 t_2 \\
 &\qquad\qquad\quad+ 63 t_1^2 t_2 - 
 7 s_1 t_2^2 - 7 s_2 t_2^2 + 63 t_1 t_2^2 + t_2^3 - 52 s_1^2 u_1 - 
 344 s_1 s_2 u_1
  \\& \qquad\qquad\quad- 52 s_2^2 u_1 + 36 s_1 t_1 u_1+ 36 s_2 t_1 u_1- 2 t_1^2 u_1 + 
 36 s_1 t_2 u_1+ 36 s_2 t_2 u_1 \\
  & \qquad\qquad\quad  + 56 t_1 t_2 u_1 - 2 t_2^2 u_1 + 43 s_1 u_1^2 + 
 43 s_2 u_1^2 
 - 7 t_1 u_1^2- 7 t_2 u_1^2 - 4 u_1^3 \\
 &  \qquad\qquad\quad - 52 s_1^2 u_2 - 
 344 s_1 s_2 u_2 - 52 s_2^2 u_2 + 36 s_1 t_1 u_2 + 36 s_2 t_1 u_2 
 - 2 t_1^2 u_2\\ 
 &\qquad\qquad\quad+ 
 36 s_1 t_2 u_2 + 36 s_2 t_2 u_2 + 56 t_1 t_2 u_2 - 2 t_2^2 u_2 + 236 s_1 u_1 u_2 + 
 236 s_2 u_1 u_2 \\
 &\qquad\qquad\quad- 74 t_1 u_1 u_2- 74 t_2 u_1 u_2
  - 72 u_1^2 u_2 + 43 s_1 u_2^2 + 
 43 s_2 u_2^2 - 7 t_1 u_2^2 \\
 &\qquad\qquad\qquad\qquad\qquad\qquad\qquad\qquad\qquad\qquad\qquad\qquad - 7 t_2 u_2^2 - 72 u_1 u_2^2 - 4 u_2^3).
\end{split}
\end{equation}
In the above computation we used tensor algebra package xAct for Mathematica.

\section{Constraint Equations}
\label{app3:constraints}

In this appendix we show how the constraint equations (\ref{constraints}) can be brought to the form (\ref{23mar5}). 

To start, we contract  (\ref{constraints}) with four momenta of fields on external lines, thus obtaining four equations
\begin{equation}
\label{25mar1}
    \lambda_n^{\alpha}\Bar{\lambda}_n^{\dot{\alpha}}\sum_{m=1}^4\frac{\partial^2 f_N}{\partial\lambda_m^{\alpha}\partial \Bar{\lambda}_m^{\dot{\alpha}}}g(\Box_p)\delta^{(4)}(p)=0, \qquad n=1,2,3,4.
\end{equation}
A straightforward computation gives 
\begin{equation}\label{25mar2}
    \begin{split}
        \frac{\partial^2 f}{\partial\lambda_1^{\alpha}\partial \Bar{\lambda}_1^{\dot{\alpha}}}=&-\lambda_{2\alpha}\Bar{\lambda}_{2\dot{\alpha}}\frac{\partial f}{\partial s_1}-\lambda_{4\alpha}\Bar{\lambda}_{4\dot{\alpha}}\frac{\partial f}{\partial t_1}-\lambda_{3\alpha}\Bar{\lambda}_{3\dot{\alpha}}\frac{\partial f}{\partial u_1}\\
        &+\lambda_{2\alpha}\Bar{\lambda}_{2\dot{\alpha}}\langle 12\rangle [12] \frac{\partial^2 f}{\partial s_1^2}+\lambda_{4\alpha}\Bar{\lambda}_{4\dot{\alpha}}\langle 14\rangle [14] \frac{\partial^2 f}{\partial t_1^2}+\lambda_{3\alpha}\Bar{\lambda}_{3\dot{\alpha}}\langle 13\rangle [13] \frac{\partial^2 f}{\partial u_1^2}\\
        &+\lambda_{4\alpha}\Bar{\lambda}_{2\dot{\alpha}}\langle 12 \rangle [14] \frac{\partial^2 f}{\partial t_1 \partial s_1}+\lambda_{3\alpha}\Bar{\lambda}_{2\dot{\alpha}}\langle 12 \rangle [13] \frac{\partial^2 f}{\partial u_1 \partial s_1}\\
        &+\lambda_{2\alpha}\Bar{\lambda}_{4\dot{\alpha}}\langle 14 \rangle [12] \frac{\partial^2 f}{\partial s_1 \partial t_1}+\lambda_{3\alpha}\Bar{\lambda}_{4\dot{\alpha}}\langle 14 \rangle [13] \frac{\partial^2 f}{\partial u_1 \partial t_1}\\
        &+\lambda_{2\alpha}\Bar{\lambda}_{3\dot{\alpha}}\langle 13 \rangle [12] \frac{\partial^2 f}{\partial s_1 \partial u_1}+\lambda_{4\alpha}\Bar{\lambda}_{3\dot{\alpha}}\langle 13 \rangle [14] \frac{\partial^2 f}{\partial t_1 \partial u_1}.
    \end{split}
\end{equation}
Other terms in the sum in (\ref{25mar1}) are obtained similarly. Focusing on equation (\ref{25mar1}) with $n=1$, we find
\begin{equation}
\label{25mar3}
    \begin{split}
        \lambda_{1}^{\alpha}\Bar{\lambda}_1^{\dot{\alpha}}\sum_{m=1}^4\frac{\partial^2 f}{\partial\lambda_m^{\alpha}\partial \Bar{\lambda}_m^{\dot{\alpha}}}&=s_1\left(\frac{\partial f}{\partial s_1}+\frac{\partial f}{\partial t_2}+\frac{\partial f}{\partial u_2}\right)+t_1\left(\frac{\partial f}{\partial t_1}+\frac{\partial f}{\partial s_2}+\frac{\partial f}{\partial u_2}\right)\\&+u_1\left(\frac{\partial f}{\partial u_1}+\frac{\partial f}{\partial s_2}+\frac{\partial f}{\partial t_2}\right)
        +s_1^2\frac{\partial^2 f}{\partial s_1^2}+t_1^2\frac{\partial^2 f}{\partial t_1^2}+u_1^2\frac{\partial^2 f}{\partial u_1^2}\\
&        +2s_1t_1\frac{\partial^2 f}{\partial s_1\partial t_1}+2s_1u_1\frac{\partial^2 f}{\partial s_1\partial u_1}+2t_1u_1\frac{\partial^2 f}{\partial t_1\partial u_1}\\
&      +(t_1+u_1)s_2\frac{\partial^2 f}{\partial s_2^2}+(s_1+u_1)t_2\frac{\partial^2 f}{\partial t_2^2}+(s_1+t_1)u_2\frac{\partial^2 f}{\partial u_2^2}\\
      & +\left(\langle 23 \rangle [24] \langle 14 \rangle [13]+\langle 24 \rangle [23] \langle 13 \rangle [14]\right)\frac{\partial^2 f}{\partial t_2 \partial u_2}\\
       & -\left(\langle 34 \rangle [23] \langle 12 \rangle [14]+\langle 23 \rangle [34] \langle 14 \rangle [12]\right)\frac{\partial^2 f}{\partial s_2 \partial t_2}\\
        &+\left(\langle 34 \rangle [24] \langle 12 \rangle [13]+\langle 24 \rangle [34] \langle 13 \rangle [12]\right)\frac{\partial^2 f}{\partial s_2 \partial u_2}.
    \end{split}
\end{equation}

First four lines of (\ref{25mar3}) are already in the desirable form: they involve only the Mandelstam variables and their derivatives. We would like to bring  the remaining lines to the same form. In flat space this can be achieved using momentum conservation. Instead, in (\ref{25mar1}) the momentum-conserving delta function appears together with the operator $g(\Box_p)$, therefore, the total momentum does not vanish identically. So, each time that we use 
\begin{equation}
\label{25mar4}
p^{\alpha\dot\alpha}\equiv \lambda^\alpha_{1}\bar\lambda^{\dot\alpha}_1+\lambda^\alpha_2\bar\lambda^{\dot\alpha}_2+\lambda^\alpha_3\bar\lambda^{\dot\alpha}_3+\lambda^\alpha_4\bar\lambda^{\dot\alpha}_4,
\end{equation}
we should keep the contribution coming from $p^{\alpha\dot\alpha}$.

To illustrate how this works in practice, we consider the contribution from the fifth line of (\ref{25mar3})
\begin{equation}
\label{25mar5}
I\equiv \frac{\partial^2 f}{\partial t_2 \partial u_2}(I_1+I_2),
\end{equation}
where
\begin{equation}
\label{25mar6}
\begin{split}
I_1\equiv \langle 23 \rangle [24] \langle 14 \rangle [13]g(\Box_p)\delta^{(4)}(p),\qquad
I_2\equiv \langle 24 \rangle [23] \langle 13 \rangle [14]g(\Box_p)\delta^{(4)}(p).
\end{split}
\end{equation}
We start by eliminating $|4\rangle |4]$ from $I_1$ by means of (\ref{25mar4})
\begin{equation}
\label{25mar7}
\begin{split}
I_1&=-\langle 23 \rangle [23] \langle 13 \rangle [13]g(\Box_p)\delta^{(4)}(p)+
\langle 23 \rangle \bar\lambda_2^{\dot\alpha} \lambda_1^\alpha [13]p_{\alpha\dot\alpha}g(\Box_p)\delta^{(4)}(p)\\
&\qquad\qquad\qquad\qquad\qquad\qquad=-u_1t_2 g(\Box_p)\delta^{(4)}(p)
-2 \langle 23 \rangle [13]\langle 1 \frac{\partial}{\partial p} 2] g'(\Box_p)\delta^{(4)}(p).
\end{split}
\end{equation}
Analogously, eliminating $|3\rangle |3]$ from $I_2$, we find
\begin{equation}
\label{25mar8}
\begin{split}
I_2=-u_2t_1 g(\Box_p)\delta^{(4)}(p)
-2 \langle 24 \rangle [14]\langle 1 \frac{\partial}{\partial p} 2] g'(\Box_p)\delta^{(4)}(p).
\end{split}
\end{equation}
So, 
\begin{equation}
\label{25mar9}
I_1+I_2=(-u_1 t_2-u_2 t_1)g(\Box_p)\delta^{(4)}(p)+I_3,
\end{equation}
where
\begin{equation}
\label{25mar10}
I_3\equiv -2 \Big( \langle 23 \rangle [13]\langle 1 \frac{\partial}{\partial p} 2]+\langle 24 \rangle [14]\langle 1 \frac{\partial}{\partial p} 2]  \Big )g'(\Box_p)\delta^{(4)}(p).
\end{equation}

Next, we proceed with $I_3$ by eliminating $|3\rangle |3]+|4\rangle |4]$
\begin{equation}
\label{25mar11}
I_3=-2 \langle 2 p 1] \langle 1 \frac{\partial}{\partial p} 2]g'(\Box_p)\delta^{(4)}(p).
\end{equation}
By commuting $p$ to the right until it multiplies the delta function and using the standard spinor algebra, we get
\begin{equation}
\label{25mar12}
I_3=16 p_1^i p_2^j \frac{\partial^2}{\partial p^i \partial p^j}g''(\Box_p)\delta^{(4)}(p)-4 s_1g''(\Box_p)\Box_p \delta^{(4)}(p)
-4 s_1g'(\Box_p)\Box_p \delta^{(4)}(p).
\end{equation}
Then, for the first two terms we use
\begin{equation}
\label{25mar13}
\begin{split}
p_1^ip_2^j \frac{\partial^2}{\partial p^i\partial p^j}g''(\Box_p)\delta^{(4)}p&=\frac{1}{4}(p_1p)(p_2p) g(\Box_p)\delta^{(4)}(p)-\frac{1}{2}p_1p_2 g'(\Box_p)\delta^{(4)}(p),\\
g''(\Box_p)\Box_p \delta^{(4)}(p)&=\frac{1}{4}p^2 g(\Box_p)\delta^{(4)}(p)-2g'(\Box_p)\delta^{(4)}(p),
\end{split}
\end{equation}
which leads to 
\begin{equation}
\label{25mar14}
I_3=(4 (p_1p)(p_2p)-s_1{p^2})g(\Box_p)
 \delta^{(4)}(p).
\end{equation}

Collecting  (\ref{25mar5}), (\ref{25mar9}) and (\ref{25mar14}), we, finally, find 
\begin{equation}
\label{25mar15}
    \begin{split}
           I= (4(p_1p)(p_2p)-s_1p^2-u_1t_2-u_2t_1)\frac{\partial^2 f}{\partial t_2 \partial u_2}g(\Box_p)\delta^{(4)}(p).
    \end{split}
\end{equation}

The last two lines in (\ref{25mar3}) can be obtained from (\ref{25mar15}) by the appropriate permutations of fields' labels
\begin{equation}
    \begin{split}
        &-\left(\langle 34 \rangle [23] \langle 12 \rangle [14]+\langle 23 \rangle [34] \langle 14 \rangle [12]\right)\frac{\partial^2 f}{\partial s_2 \partial t_2}g(\Box_p)\delta^{(4)}(p)\\
        &\hspace{2.5cm}=(4(p_1p)(p_3p)-u_1p^2-t_1s_2-t_2s_1)\frac{\partial^2 f}{\partial t_2 \partial s_2}g(\Box_p)\delta^{(4)}(p),
    \end{split}
\end{equation}
\begin{equation}
    \begin{split}
        &\left(\langle 34 \rangle [24] \langle 12 \rangle [13]+\langle 24 \rangle [34] \langle 13 \rangle [12]\right)\frac{\partial^2 f}{\partial s_2 \partial u_2}g(\Box_p)\delta^{(4)}(p)\\
        &\hspace{2.5cm}=(4(p_1p)(p_4p)-t_1p^2-u_1s_2-u_2s_1)\frac{\partial^2 f}{\partial u_2 \partial s_2}g(\Box_p)\delta^{(4)}(p).
    \end{split}
\end{equation}

Combining all the contributions to (\ref{25mar1}) with $n=1$ and equating the prefactor of $g(\Box_p)\delta(p)$ to zero, we find  
the first equation in (\ref{23mar5}). Other equations, again, can be obtained by permuting fields' labels.

\bibliography{shv3}
\bibliographystyle{JHEP}

\end{document}